\documentclass[sn-mathphys-num]{sn-jnl}

\usepackage{graphicx}%
\usepackage{multirow}%
\usepackage{amsmath,amssymb,amsfonts}%
\usepackage{amsthm}%
\usepackage{mathrsfs}%
\usepackage[title]{appendix}%
\usepackage{xcolor}%
\usepackage{textcomp}%
\usepackage{manyfoot}%
\usepackage{booktabs}%
\usepackage{algorithm}%
\usepackage{algorithmicx}%
\usepackage{algpseudocode}%
\usepackage{listings}%

\theoremstyle{thmstyleone}%
\theoremstyle{thmstyletwo}%

\theoremstyle{thmstylethree}%

\begin{document}

\title[Article Title]{Superconductivity in Janus IV-B transition metal chalcogenide hydrides}


\author[1]{\fnm{Jakkapat} \sur{Seeyangnok}}\email{jakkapatjtp@gmail.com}

\author*[1]{\fnm{Udomsilp} \sur{Pinsook}}\email{Udomsilp.P@Chula.ac.th}

\author*[2]{\fnm{Graeme John} \sur{Ackland}}\email{gjackland@ed.ac.uk}

\affil[1]{\orgdiv{Department of Physics, Faculty of Science}, \orgname{Chulalongkorn University}, \orgaddress{\street{254 Phaya Thai Rd}, \city{Bangkok}, \postcode{10330}, \country{Thailand}}}

\affil[2]{\orgdiv{Centre for Science at Extreme Conditions, School of Physics and Astronomy}, \orgname{University of Edinburgh},  \city{Edinburgh}, \postcode{EH9 3FD}, \state{Scotland}, \country{United Kingdom}}


\abstract{Two-dimensional  Janus transition-metal chalcogenide -hydrides (2D-JTMCs) feature a three layered structure, with a central layer of transition metal atoms, with chalcogenides below and hydogens above.  This asymmetry endows 2D-JTMCs with unique and tunable electronic, optical, and mechanical properties. In this paper, we systematically investigate two-dimensional hexagonal 2H and 1T Janus transition metal chalcogenide (JTMC) hydrides (MXH), with $M = \text{Ti, Zr, Hf}$, and $X = \text{S, Se, Te}$. These JTMCs exhibit metallic behavior with single band crossings at the Fermi level, primarily dominated by transition metal $d$ orbitals and phonon stability, evidenced by DFPT calculations. We estimate $T_{c}$ using the Allen-Dynes formula and the closing superconducting gap by solving anisotropic gap equations of the Migdal-Eliashberg equations. Calculated $T_{c}$ values for these materials range from 10 K to 35 K.}




\maketitle

	\section{Introduction}
    Two-dimensional (2D) materials have garnered significant interest due to their diverse physical properties and potential applications in nanomaterials and condensed matter physics since the groundbreaking discovery of graphene in 2004 \cite{novoselov2004electric}. Phonon-mediated superconductors \cite{bardeen1957microscopic} are explained by electron-phonon interaction based on Migdal-Eliashberg (ME) theory \cite{frohlich1950theory,migdal1958interaction,eliashberg1960interactions,nambu1960quasi}. Concurrently, superconducting hydrides have achieved remarkable critical temperatures, inspired by Ashcroft's predictions of room temperature superconductivity in metallic hydrogen at extreme pressures \cite{ashcroft1968metallic,ashcroft2004hydrogen} due to the light mass of hydrogen which leads to high-frequency phonon modes. This combination of materials is particularly appealing, leading to investigations into new forms of 2D hydrides that might exhibit high-temperature superconductivity without the need for extreme pressures, alongside intriguing physical properties from their 2D structure. For instance, hydrogenated graphene has been proposed as a conventional superconductor with a $T_c$ above 90K \cite{sofo2007graphane}. Further examples include hydrogenated MgB$_2$ monolayers (ML) with a predicted $T_c$ of 67K \cite{savini2010first}, and hydrogenated HPC$_3$ with a predicted $T_c$ of 31K \cite{li2022phonon}.

    2D transition metal dichalcogenides (2D-TMDs) comprise a close-packed monolayer of transition metal atoms sandwiched between layers of chalcogenide atoms. The positions of the chalcogenides give rise to so-called 1T and 2H structures (Fig. \ref{fig:structures}, the relative stability of which can be tuned by strain provided by a substrate.
    
    These materials have been intensely studied due to their flat electronic band structures \cite{zhang2020flat,vitale2021flat,kuang2022flat,rademaker2022spin,huang2023recent,hinlopen2024lifshitz}. In contrast, 2D Janus transition-metal dichalcogenides (2D-JTMDs) feature an out-of-plane asymmetric layered structure, with different chalcogenide atoms on each facet. This asymmetry endows 2D-JTMDs with unique and tunable electronic, optical, and mechanical properties \cite{tang20222d,varjovi2021janus,zhang2022janus,angeli2022twistronics,maghirang2019predicting,he2018two,yeh2020computational,yin2021recent,li2023structure}. Although 2D-JTMDs do not occur naturally, they have been synthesized, starting with Janus graphene in 2013 \cite{zhang2013janus}. Numerous 2D-JTMDs have since been created, such as MoSSe \cite{trivedi2020room,lu2017janus}, WSSe \cite{trivedi2020room}, and PtSSe \cite{sant2020synthesis}. These materials are typically fabricated using selective epitaxy atomic replacement (SEAR) processes with hydrogen (H$_2$) plasma \cite{trivedi2020room,tang20222d}. Recently, a Janus 2H-MoSH ML was synthesized using the SEAR method by substituting the top S atoms with hydrogen atoms \cite{lu2017janus}. Theoretically, Janus MoSH ML has been predicted to be a superconductor with $T_c$ = 26.81K for the 2H phase \cite{liu2022two,ku2023ab}. Furthermore, the 2H-WSH phase has been suggested to be dynamically stable with a $T_c$ above 12K \cite{seeyangnok2024superconductivity}. This phase stability and superconductivity have been corroborated by subsequent independent studies \cite{gan2024hydrogenation,fu2024superconductivity}. 

    Moreover, there have been many investigations of physical properties of group IV transition-metal dichalcogenides (MX$_2$, M = Ti,Zr,Hf; X = S,Se,Te), for incomplete list\cite{joseph2023review,lasek2021synthesis,mattinen2019atomic,zhang2016systematic,toh2016catalytic,xie2015two}. We  expect that SEAR will be able to produce JTMC hydrides, for which superconductivity has been predicted in 2H-TiSH, 2H-TiSeH, 1T-TiSH, 1T-TiSeH, 1T-TiTeH, and 1T-ZrTeH 
    in the 10-30K range\cite{li2024machine,ul2024superconductivity}. Therefore, in this paper, we systematically study Janus 2D chalcogenide (JTMCs) hydrides in both 2H and 1T phases  with group IV transition metals Ti, Zr, Hf and chalcogens {S, Se, Te}. We commence with a discussion of the electronic structure, followed by an examination of phase stability, thermal effect of phase stability and subsequently delve into the exploration of superconductivity.

\section{Results and Discussion}
    	\subsection{Crystal structures}

    \begin{figure}[h]
		\centering  
    	\includegraphics[width=12cm]{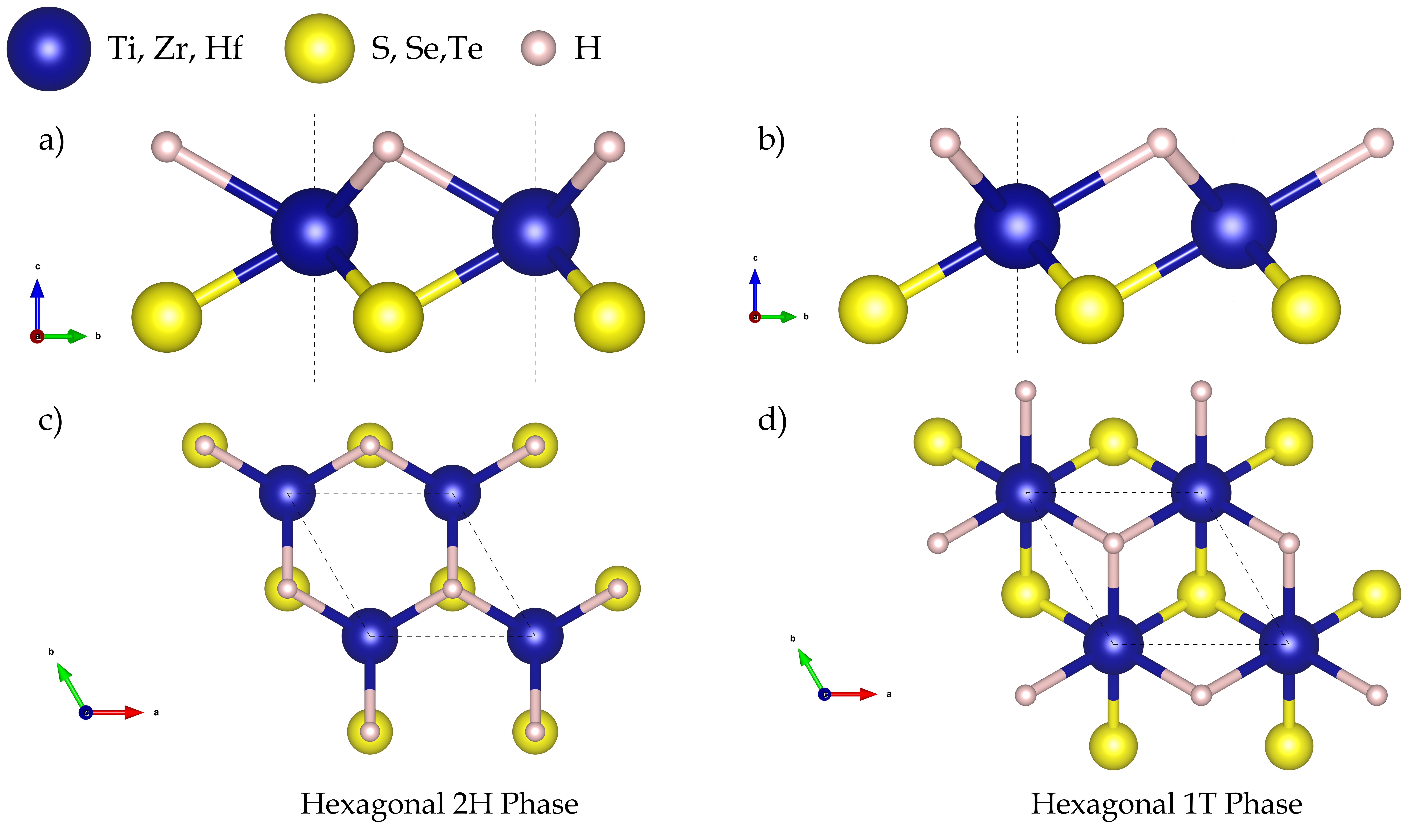}
		\caption{(a, c) and (b, d), show side-view and top-view crystal of two-dimensional (2D) hexagonal 2H and 1T Janus IV-TMCs hydrides where the blue ball represents Ti, Zr and Hf and the yellow ball represents S, Se, and Te, and the hydrogen is denoted by the pink ball, respectively.}
		\label{fig:structures}
	\end{figure} 

    The periodically repeated 2H and 1T phase of structure of our 2D JTMCs $MXH$ (M = Ti, Zr, Hf, and X = S, Se, Te) belongs to the 3D trigonal space group P3m1 (No.156) as shown in Fig.~\ref{fig:structures}. In the $x-y$ coordinates, these have a hexagonal arrangement  lattice constant $a$; Transitional metals ($M$) sit at (0,0) positions while chalcogenide groups ($X$) sit at (1/3,2/3). The hydrogen is at (1/3,2/3) for the 2H phase and (2/3,1/3) for 1T the phase: ABA and ABC stacking respectively. Along the $z$ axis, the structure  can be defined by two variables: distance between $M$- and H-layers, and the distance between $M$- and $X$-layers, denoted by $h_{\text{H}}$ and $h_{X}$, respectively. These cell variables are not fixed by symmetry and the values we obtained are given in Table~\ref{tab:lattice-const}. There are 18 possible 
     MXH combinations. However, we found that 2H-TiTeH, 1T-ZrSH, 2H-ZrTeH, 1T-HfSH, 1T-HfSeH, and both HfTeH structures are totally unstable against flexural acoustic phonons, see detail in Table~\ref{tab:energy-structure}. Although for applications this could be compensated by a substrate, we examined superconductivity in only the 11 stable combinations  Table~\ref{tab:lattice-const}. 
    
    \begin{table}[h!]
    \centering
	   \begin{tabular}{cccc}
    \hline
		JTMCs & $a$ (\AA) & $h_{\text{H}}$ (\AA) & $h_{X}$ (\AA) \\
  \hline
        2H-TiSH & 3.15 & 1.06 & 1.50 \\
        2H-TiSeH & 3.23 & 0.99 & 1.66 \\
        2H-ZrSH & 3.39 & 1.11 & 1.56 \\
        2H-ZrSeH & 3.45 & 1.05 & 1.72 \\
        2H-HfSH & 3.34 & 1.11 & 1.57 \\
        2H-HfSeH & 3.41 & 1.05 & 1.72 \\
        \hline
        1T-TiSH & 3.21 & 0.87 & 1.46 \\
        1T-TiSeH & 3.27 & 0.80 & 1.63 \\
        1T-TiTeH & 3.37 & 0.67 & 1.92 \\
        1T-ZrSeH & 3.50 & 0.83 & 1.70 \\
        1T-ZrTeH & 3.59 & 0.69 & 1.96 \\
        \hline
	   \end{tabular}
    \caption{Calculated lattice constants $a$, the distance between $M$- and H-layers $h_{\text{H}}$, and  $M$- and $X$-layers $h_{X}$.}
	\label{tab:lattice-const}
	\end{table}	

    The general trends are as follows; \begin{itemize}
        
   \item For fixed X, $a$ increases as M = Ti $\rightarrow$ Zr $\rightarrow$ Hf. \item For fixed M, $h_{\text{H}}$ decreases and $h_{\text{H}}$ increases  as X = S $\rightarrow$ Se $\rightarrow$ Te. \item In a given compound, $h_{\text{H}}$ is larger in 1T than 2H.
 \end{itemize}

 A high density of states near the Fermi surface is often associated with an instability in the metallic state\cite{ackland2004origin}, be it structural, superconducting or magentic.
 For these structures, we have investigated magnetic phases, including ferromagnetism and antiferromagnetism (G-type antiferromagnetism (GAF), and C-type antiferromagnetism (CAF). We found that the ferromagnetic phase is energetically favorable for the 2H phase of TiSH, TiSeH, ZrSH, and ZrSeH, while the non-magnetic phase is the most energetically favorable for the 2H phase of HfSH and HfSeH. For the 1T phases, TiSH, TiSeH, TiTeH, ZrSeH, and ZrTeH exhibit a favorable ferromagnetic phase. All information is shown in Table~\ref{tab:phase-energies}. These findings provide more information that has not been considered in other studies \cite{ul2024superconductivity,li2024machine}: we have checked the stability of other JTMCs, finding that e.g. Mo an W are stable, whereas putative JTMCs involving Cr or Mn are highly unstable against magnetic states.

    \begin{table}[h!]
    \centering
	   \begin{tabular}{cccc}
        \hline
		JTMCs  & FM & GAF & CAF \\
	   \hline
        2H-TiSH  & -23.44936 & 0.60404 & 0.60373  \\
        2H-TiSeH  & -36.07079 & 0.44364 & 0.44347 \\
        2H-ZrSH  & -19.79324 & 0.79168 & 0.79141 \\
        2H-ZrSeH  & -27.45720 & 0.78371 & 0.78659 \\
        2H-HfSH  & 0.28324 & 0.29277 & 0.28157 \\
        2H-HfSeH  & 0.17320 & 0.27270 & 0.27300 \\
        \hline
        1T-TiSH  & -20.18296 & 0.00233 & 0.00179 \\
        1T-TiSeH  & -23.64239 & 0.01457 & 0.01413 \\
        1T-TiTeH  & -23.87525 & 0.00440 & 0.00092 \\
        1T-ZrSeH  & -13.31320 & 0.08259 & 0.08203 \\
        1T-ZrTeH  & -10.23288 & 0.00252 & 0.00145 \\
            \hline
	   \end{tabular}
    \caption{Calculated relative energies of magnetic phases with respect to their non-magnetic phase ($E_{M}-E_{2H}$) of FM, GAF, and CAF in the unit of meV.}
	\label{tab:phase-energies}
	\end{table}

    \subsection{Electronic structures}
 
\begin{figure}[h!]
		\centering  
    	\includegraphics[width=12cm]{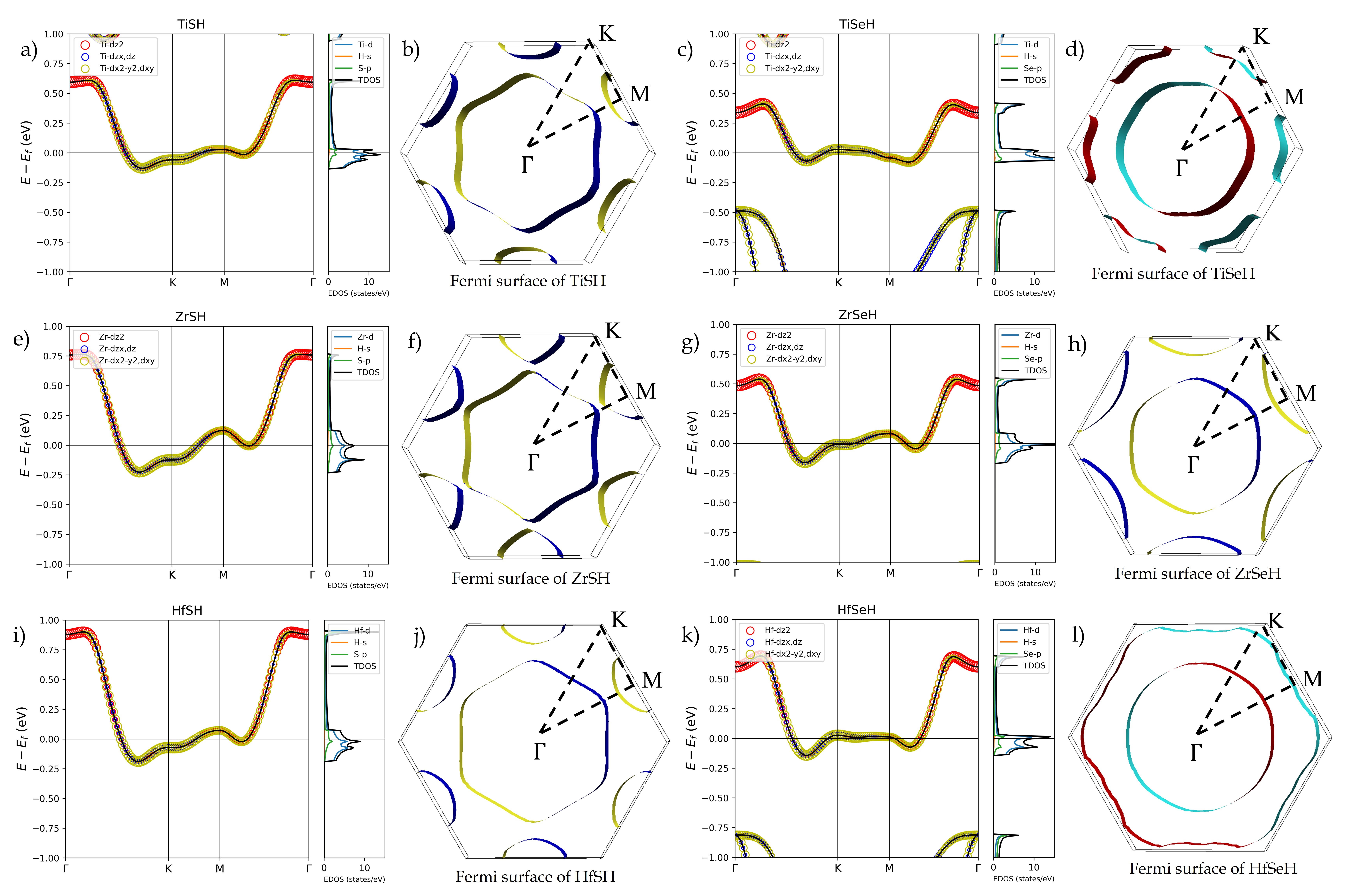}
		\caption{Electronic properties of 2H compounds close to the Fermi level (a-l) figures show the orbital-resolved electronic band structures, the electronic density of states, the orbital projected density of states and the Fermi surface of 2D 2H-JTMC (2H-MXH), with $M = \text{Ti, Zr, Hf}$ and $X = \text{S, Se}$. All 2H Tellurium-based structures are dynamically unstable and are not shown.}
		\label{fig:mxh-electronics}
	\end{figure}

    The Kohn-Sham electronic properties, including the orbital-resolved electronic band structures, the electronic density of states, the orbital projected density of states and the Fermi surface in the Brillouin zone are shown in Fig.~\ref{fig:mxh-electronics}. In general, these materials share many similarities in their electronic band structures for relative energies ($E-E_{f}$) near the Fermi level. These JTMCs exhibit metallic behavior.  For 2H phase, the bands at the Fermi level are predominantly of TM $E'(d_{xy}, d_{x^2-y^2})$ orbital character, but hybridized with all other orbitals.  

    From Figs.~\ref{fig:mxh-electronics} (a) and (c), it is evident that a single $d$-band is responsible for crossing at the Fermi level, leading to the metallic properties of TiSH and TiSeH. Despite the similarity of TiSH and TiSeH, different topology can be seen at the $M$ and $K$ points.  In TiSH, the band intersects the Fermi surface around the $M$ point, whereas in TiSeH, the intersection occurs around the $K$ point. This produces a pocket of holes and a flattened section of band structure around the $K$ point for TiSeH, or around the $M$ point in TiSH.  These flat dispersions are associated with a high density of states near the Fermi level, as shown in the projected electronic density of states. They suggest van Hove singularities (vHS) near the Fermi level.

    The electronic structure of ZrSH and ZrSeH is shown in Fig.~\ref{fig:mxh-electronics} (e-h). The electronic structure of both ZrSH and ZrSeH is very similar to that of TiSH, characterized by a  Fermi surface enclosing hole-pockets around the $\Gamma$ point and the $M$ point. These similarities result from the same intersection of the bands along the symmetry points $\Gamma \rightarrow K \rightarrow M \rightarrow \Gamma$. However, despite having a flattened band around the $K$ point for ZrSeH, ZrSH does not exhibit this feature, resulting in a lower electronic density of states, as shown in Fig.~\ref{fig:mxh-electronics} (e).

    Finally, the electronic structures of HfSH and HfSeH  (Figs.~\ref{fig:mxh-electronics} (i)-(l))  are similar to those of TiSH, ZrSH, and ZrSeH. A very interesting behavior is observed in HfSeH, where a flattened electron band along the $K$ to $M$ direction results in a van Hove singularity (vHS) of high electronic density of states near the Fermi level. This is associated with the vHS, as shown in Fig.~\ref{fig:mxh-electronics} (k).

    \begin{figure}[h!]
		\centering  
    	\includegraphics[width=12cm]{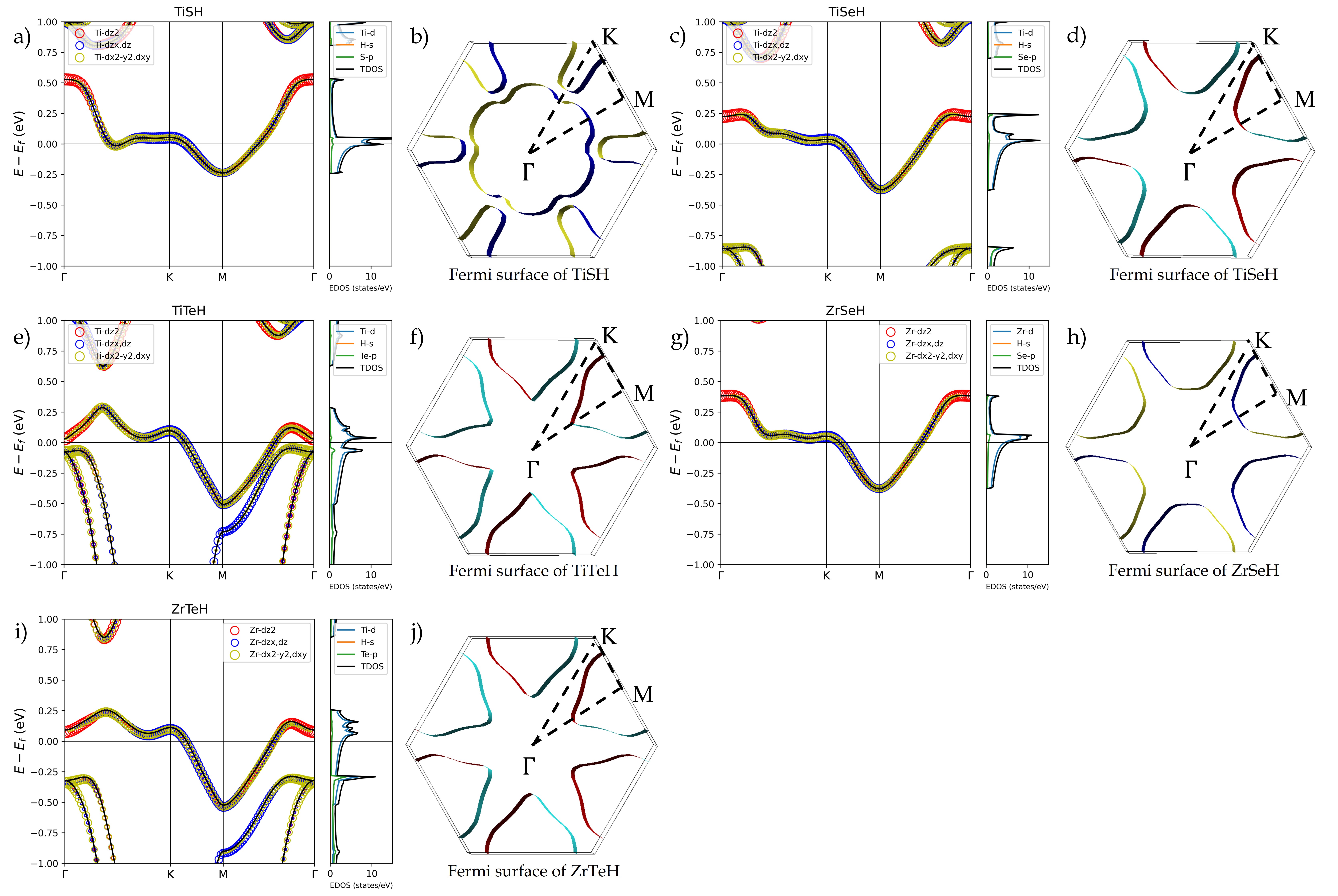}
		\caption{ Electronic properties of 1T compounds (a-j) show the orbital-resolved electronic band structures, the electronic density of states, the orbital projected density of states and the Fermi surface of 2D 1T-JTMC (1T-MXH) of TiSH, TiSeH, TiTeH, ZrSeH, and ZrTeH. It is worth noting that 1T-ZrSH and all HfXH are dynamically unstable.}
		\label{fig:1tmxh-electronics}
    \end{figure}

     For the 1T phase of JTMCs, the electronic structure of dynamically stable phonons is shown in Figure~\ref{fig:1tmxh-electronics}. We found that 1T-ZrSH and all HfXH have instabilities against flexural acoustic (ZA) distortions, see detail in Table~\ref{tab:energy-structure}. We chose not to report these compounds further. From Figure~\ref{fig:1tmxh-electronics}, the stable JTMCs exhibit metallic behavior due to single band crossings at the Fermi level, which are  predominantly both $E'(d_{xy}, d_{x^2-y^2})$ and $E''(d_{yz}, d_{xz})$. 
     The electronic structure of 1T-JTMCs is similar to that of 2H JTMCs, except that the electron pocket is at the M-point. From Figs.~\ref{fig:1tmxh-electronics},  which leads to a van Hove singularity (vHS) with a high electronic density of states. 
     TiSH is topologically different in that band dips below the Fermi level between $\Gamma$ and $K$, so that the M-pockets are joined.
    
   The electronic dispersion has significantly lower energy around the M-point of the 1T structure, compared with that of 2H. This indicates that the H atom might form a stronger bonding in the 1T-structure than in 2H. This should explain why $h_{\text{H}}$ of the same compound decreases as 2H $\rightarrow$ 1T. Also, we will see later that the in-plane hydrogen phonon mode has  significantly higher frequencies in the 1T-structure than in 2H.

   Figures~\ref{fig:mxh-electronics} and Figures~\ref{fig:1tmxh-electronics} demonstrate that these materials exhibit a very high electronic density of states at the Fermi level, which typically favors electronic instability. In addition to investigation of the electronic instability associated with Cooper pairs in superconducting states, it is shown that the 2H phase of TiSH, TiSeH, ZrSH, and ZrSeH has the potential to become ferromagnetic, while the non-magnetic 2H phase of HfSH and HfSeH remains the most energetically favorable. In the 1T phase, TiSH, TiSeH, TiTeH, ZrSeH, and ZrTeH also have the potential to become ferromagnetic. However, the magnetic phase energies are approximately 10-30 meV lower than the normal metallic phase, as shown in Table~\ref{tab:phase-energies}. This analysis does not yet account for energy gain from the Cooper pair instability from the metallic state, which could result in an even lower energy for the superconducting phase. 

   	\subsection{Phonon properties.}
\begin{figure}[h]
		\centering  
    	\includegraphics[width=12cm]{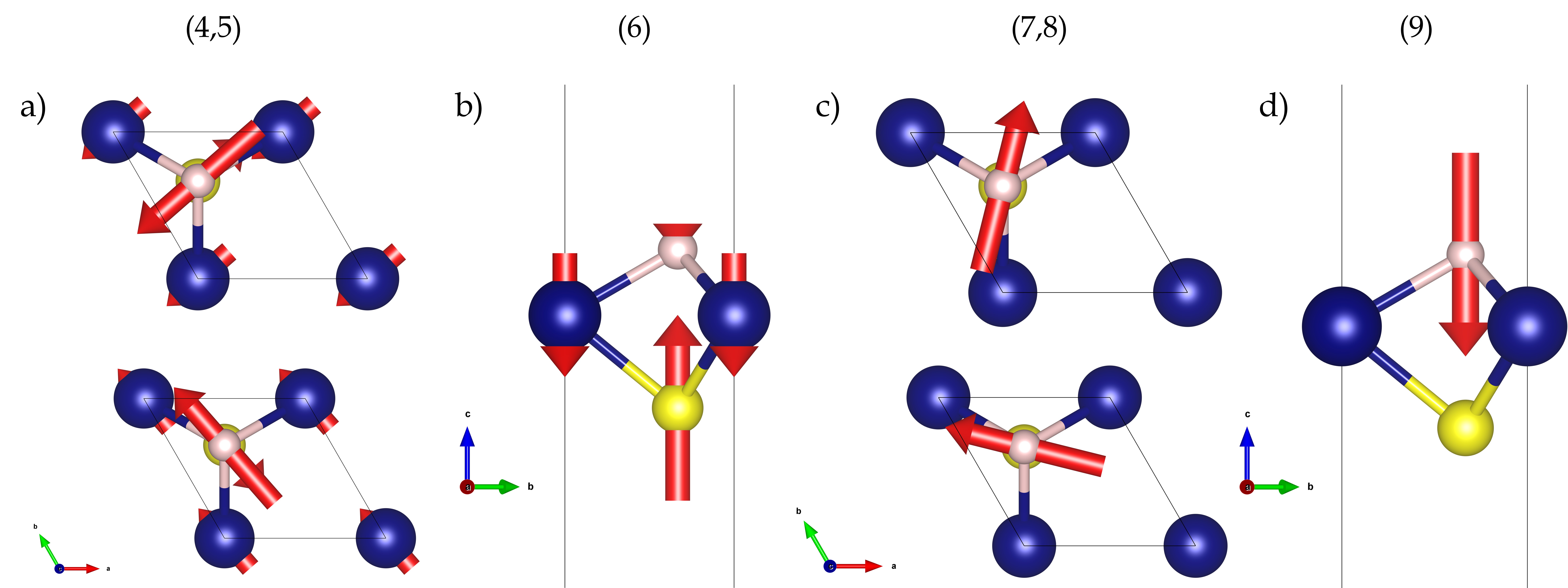}
		\caption{(a-d) shows the six optical vibrational modes of Janus transition-metal monolayer at $\Gamma$ point in the Brillouin zone. The color of spheres is the same as Fig.~\ref{fig:structures}: red=M, yellow=X, white=H. Arrows indicate the direction and magnitude of the eigenvector. The frequencies are given in Table~\ref{tab:vibrational-modes}.}
		\label{fig:normal-modes}
    \end{figure}

    JTMCs have the P3m1 trigonal space group, and three-atoms primitive units cell. Therefore, the phonons have symmetry modes of the $C_{3v} (3m)$ point group at the $\Gamma$ point, and consist of three acoustic branches and six optical branches. The three acoustic branches include an in-plane longitudinal mode (LA), an in-plane transverse mode (TA), and an out-of-plane flexural mode (ZA). These are labeled as band 1-3. There are 6 optical bands, labeled as band 4-9. Their corresponding eigenvectors at $\Gamma$ point are shown in Figures~\ref{fig:normal-modes}: the low mass of hydrogen means that three bands can readily be identified with high frequency hydrogen motion.
 
            \begin{table}[h]
	\begin{tabular}{ccccccccc}
        \hline
		Modes & Sym. & Eigenvec. & 2H-TiSH & 2H-TiSeH & 2H-ZrSH & 2H-ZrSeH & 2H-HfSH & 2H-HfSeH \\
        \hline
		4,5 & E & \text{IP $X$, H}      & 25.4 & 19.4 & 20.0 & 15.6 & 18.9 & 13.5\\
		6 & $A_{1}$ & \text{OP $X$, H} & 52.7 & 41.0 & 46.5 & 33.5 & 45.5 & 30.6\\
		7,8 & E & \text{IP H}           & 83.0 & 81.3 & 86.5 & 86.1 & 86.0 & 84.5 \\
		9 & $A_{1}$ & \text{OP H}      & 128.0 & 121.4 & 123.0 & 116.9 & 127.2 & 120.9 \\
        \hline
        Modes & Sym. & Eigenvec. & 1T-TiSH & 1T-TiSeH & 1T-TiTeH & 1T-ZrSeH & 1T-ZrTeH  \\
        \hline
		4,5 & E & \text{IP $X$, H}      & 31.0 & 23.5 & 17.2 & 19.0 & 14.6 \\
		6 & $A_{1}$ & \text{OP $X$, H} & 51.3 & 39.2 & 28.5 & 32.9 & 26.6 &\\
        7,8 & E & \text{IP H}          & 114.4 & 113.7 & 111.8 & 115.7 & 113.7  \\
		9 & $A_{1}$ & \text{OP H}       & 111.7 & 105.3 & 96.9 & 95.8 & 86.5 & \\
        \hline
		\end{tabular}
    \caption{The table shows mode number, subgroup symmetry (Sym.), vibrational eigenvectors (Eigenvec.), and the frequencies of the six optical modes at the $\Gamma$ point of 2D JTMC (MXH), with $M = \text{Ti, Zr, Hf}$ and $X = \text{S, Se, Te}$. The unit of the phonon frequencies is meV. The eigenvectors are labelled X and H depending on whether chalcogenide or hydrogen atoms are involved. All modes are infrared and Raman actives. IP denotes an in-plane vibration, and OP denotes an out-of-plane vibration.}
		\label{tab:vibrational-modes}
	\end{table}
 
\begin{figure}[h]
		\centering  
    	\includegraphics[width=12cm]{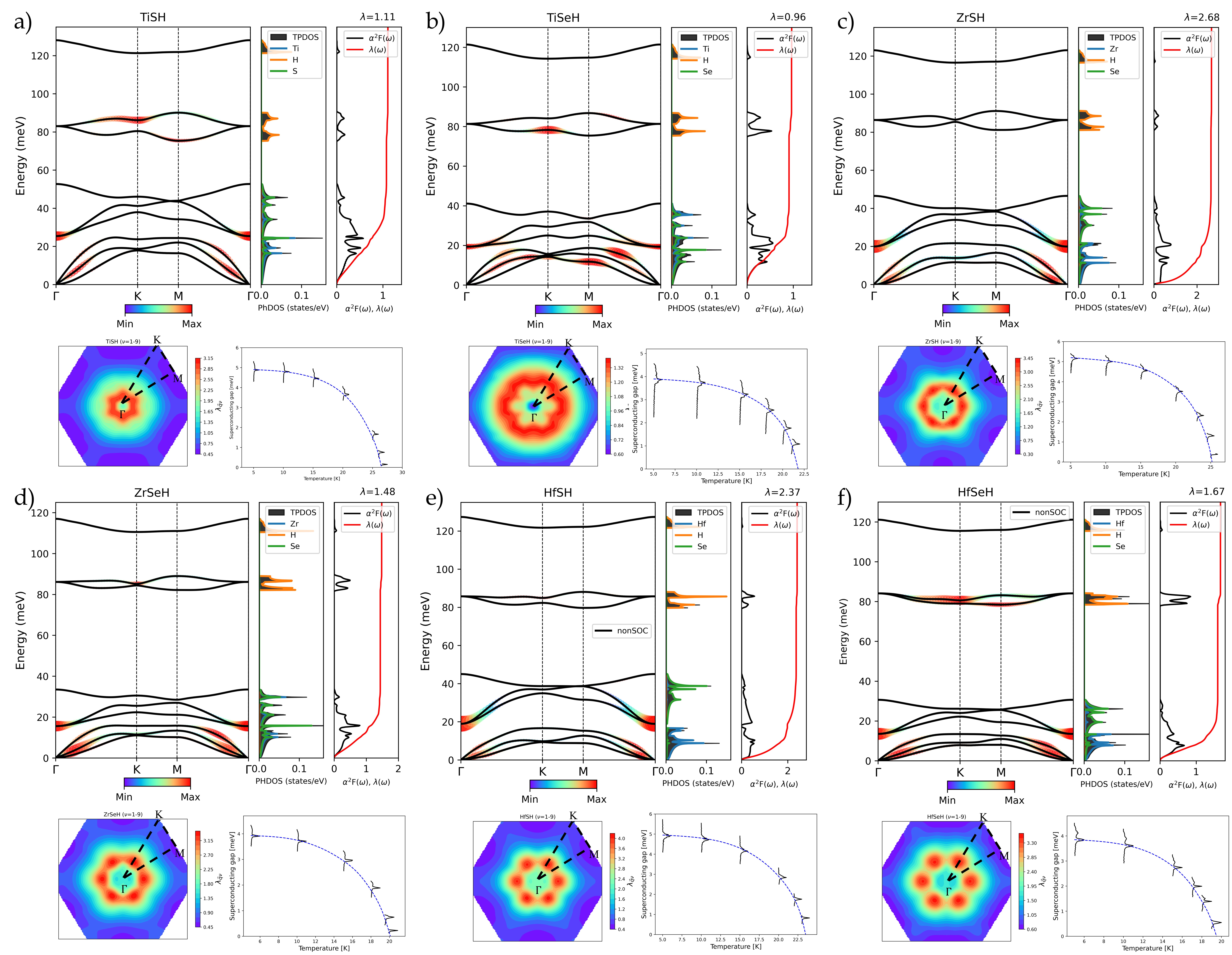}
		\caption{(a-f) show the phonon properties and electron-phonon interaction of 2D 2H-JTMC (2H-MXH), with $M=\text{Ti, Zr, Hf}$ and $X=\text{S, Se}$. It consists of the the weighted electron-phonon coupling (EPC) phonon dispersion, phonon density of states, the isotropic Eliashberg spectral function $\alpha^2 F (\omega)$ and $\lambda (\omega)$, and the coutour of EPC in the Brillioun zone, and the superconducting gap.}
		\label{fig:elph-tixh}
	\end{figure}

   The phonon spectra for the dynamically stable materials, are shown in Fig.\ref{fig:elph-tixh} (a)-(f).
    The lowest frequency optical bands labelled 4-5 are  in-plane vibrations and the band 6  is the out-of-plane vibration of chalcogenides (\(X = \text{S, Se, Te}\)) and hydrogen, as shown in (a) and (b) of Figures~\ref{fig:normal-modes}, respectively. The second set of optical bands, referred to as bands 7-9, includes bands 7-8 arising from the in-plane vibration of hydrogen and band 9 resulting from the out-of-plane vibration of hydrogen as shown in (c) and (d) of Figures~\ref{fig:normal-modes}, respectively. The absence of 3D inversion symmetry makes all $\Gamma$-point modes Raman and IR active (Table~\ref{tab:vibrational-modes}).  It is also worth noting that, in the 2H phase, all tellurium compounds are unstable.

    The vibrational frequencies of the hydrogen-based modes 7, 8, and 9 are similar across our JTMCs. However, in the vibrational frequencies of modes 4, 5, and 6, there is a difference between the transition metal sulfur and selenium hydrides. If this were just a mass effect, the harmonic approximation, suggests that a factor is of the square root of the mass ratio between the two, $\sqrt{2.46} \approx 1.57$. Due to the coupled vibration of hydrogen, the actual ratio could be slightly different from this number, depending on the bonding among transition metals, chalcogenides, and hydrogen, which appears to be about 1.3-1.4. This closely aligns with the predictions from lattice dynamics theory within the harmonic approximation.

        \begin{figure}[h]
		\centering  
    	\includegraphics[width=12cm]{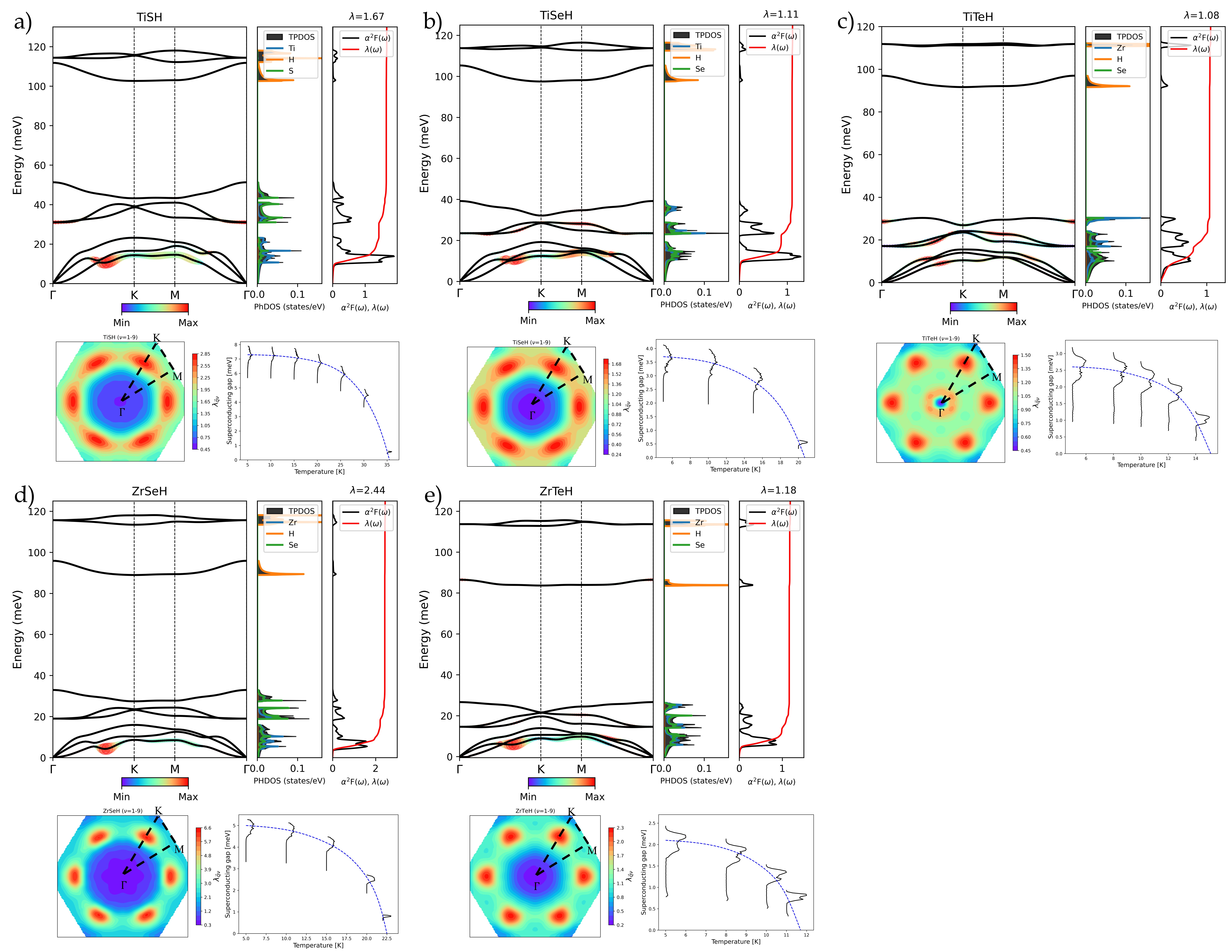}
		\caption{(a-e) show the phonon properties and electron-phonon interaction of 1T phase of TiSH, TiSeH, TiTeH, ZrSeH, and ZrTeH. It consists of the the weighted electron-phonon coupling (EPC) phonon dispersion, phonon density of states, the isotropic Eliashberg spectral function $\alpha^2 F (\omega)$ and $\lambda (\omega)$, and the contour of EPC in the Brillioun zone, and the superconducting gap.}
		\label{fig:1telph-tixh}
	\end{figure}

    Phonon spectra for stable 1T structure are shown in Figures~\ref{fig:1telph-tixh} (a)-(e). The vibrational properties of the 1T phase of these JTMCs are similar to those of the 2H phase, as illustrated in Figures~\ref{fig:normal-modes}. The only significant difference is that the highest energy mode of vibration for the 1T phase results from the out-of-plane vibration of hydrogen (modes 7 and 8) rather than the in-plane vibration of hydrogen (mode 9), as occurs in the 2H phase. The energies of the vibrational modes of the dynamically stable 1T phase are also summarized in Table~\ref{tab:vibrational-modes}. According to the theory of lattice dynamics with the harmonic approximation, we expect a factor of the square root of the mass ratio between Se and Te, $\sqrt{1.61} \approx 1.27$. The actual ratio of mode 4, 5 and mode 6 of chalcogenide frequencies between TiSeH and TiTeH, and ZrSeH and ZrTeH from the DFPT calculation agrees very well with this ratio from lattice dynamics theory within the harmonic approximation.

    For the unstable phonons of JTMCs, as listed in Table~\ref{tab:energy-structure}, the following observations were made: In the 2H phase, TiTeH and HfTeH exhibit negative frequencies around the $K$ point from the ZA flexural phonon mode, while ZrTeH shows negative ZA-phonon frequencies between the $\Gamma$ to $K$. In the 1T phase, ZrSH and HfSeH display negative ZA frequencies between $\Gamma$ and  $K$, while HfSH also exhibits additional negative ZA-mode frequencies around the $M$ point. Additionally, HfTeH shows negative  frequencies for both the ZA and TA  modes across the Brillouin zone of wavevector $q$. The soft (ZA) flexural phonon mode  implies  strong electron-phonon coupling as shown in Figures~\ref{fig:1telph-tixh}.

   Thus the general trends of the $\Gamma$ point phonons,   ignoring the unstable compounds,
      are as follows; 
   \begin{itemize}

   \item For fixed M, the frequencies of bands 4-6 decrease as X = S $\rightarrow$ Se $\rightarrow$ Te. \item For a fixed X, the frequencies of bands 4-6 decrease as M = Ti $\rightarrow$ Zr $\rightarrow$ Hf. \item For a fixed M, the frequencies of bands 7-9 decrease as X = S $\rightarrow$ Se $\rightarrow$ Te. \item For a fixed X, the frequencies of the band 7, 8 are highest for Zr, while  band 9 is lowest for Zr.   
   \end{itemize}

    \subsection{Thermal stability of hexagonal 2H Janus phase.}
    
    \begin{figure}[h]
		\centering  
    	\includegraphics[width=12cm]{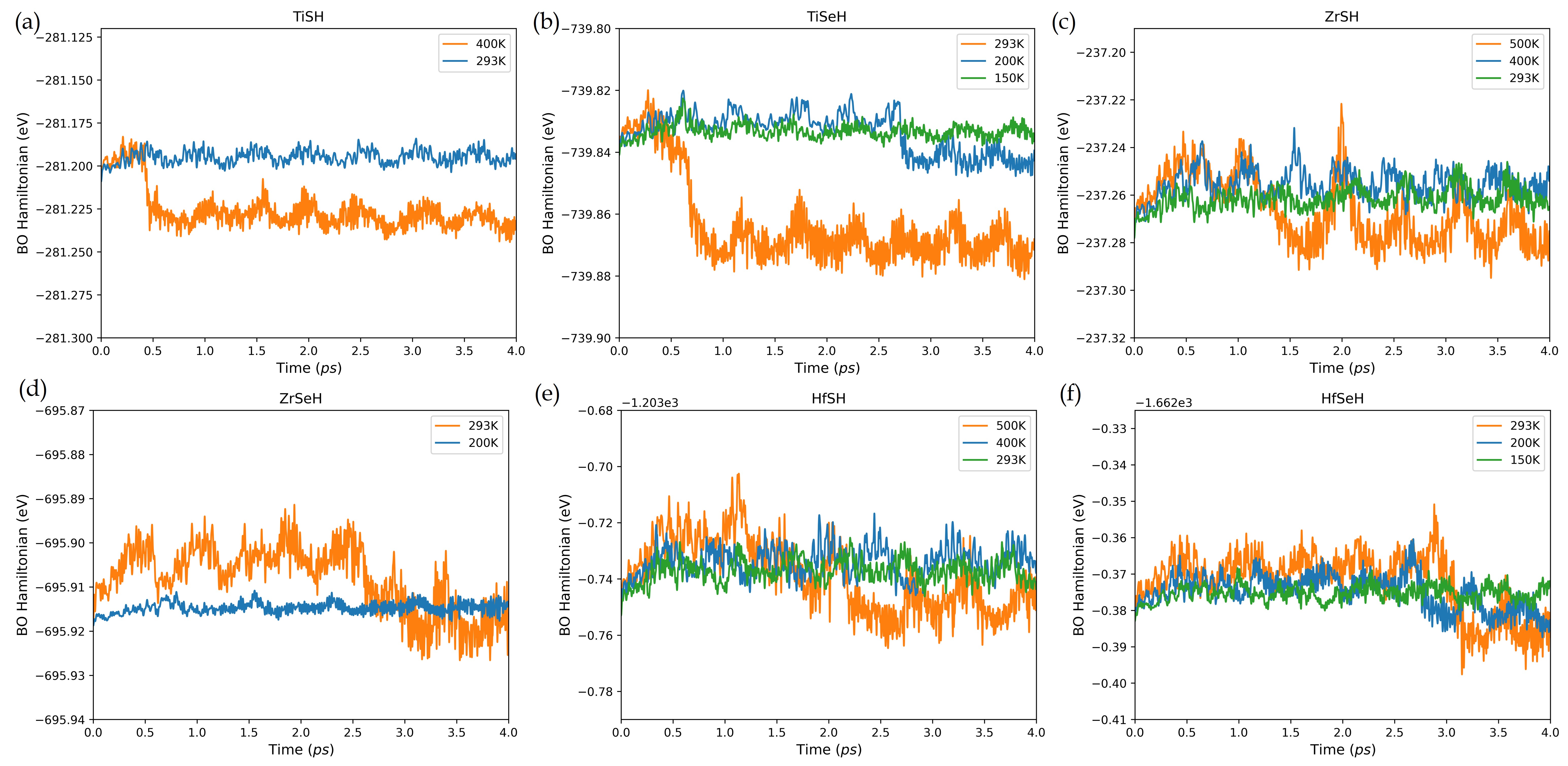}
		\caption{(a-f) show the Born-Oppenheimer Hamiltonain from the molecular dynamics (MD) of 2D 2H-JTMC (2H-MXH), with $M=\text{Ti, Zr, Hf}$ and $X=\text{S, Se}$ computed from CASTEP \cite{CASTEP}.The 2$\times$2$\times$1 supercell MD shows BO Hamiltonain energy different temperatures.}
		\label{fig:MD-analysis}
	\end{figure}

\begin{figure}[h]
		\centering  
    	\includegraphics[width=12cm]{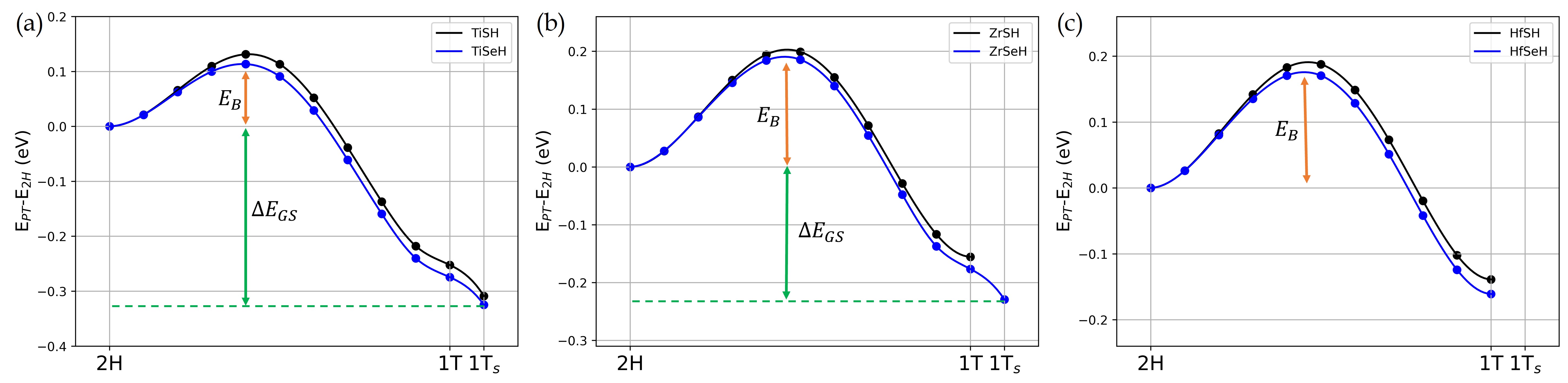}
		\caption{Figures (a-c) show the energy barriers between the stable 2H phase, the non-optimized 1T phase (with lattice constants of the 2H phase), and the stable phonon of the optimized 1T phase (referred to as $1T_{s}$) of JTMCs $MXH$ where $M = \text{Ti, Zr, Hf}$ and $X = \text{S, Se}$. For JTMCs where $1T_{s}$ is not indicated, the phonons of the optimized 1T phase are not stable. The transition from the 2H to the 1T structures is simulated by a series of step-wise displacements of the hydrogen atom from $(1/3, 2/3)$ to $(2/3, 1/3)$ in the 2D coordinates.}
		\label{fig:energy-barrier}
	\end{figure}

    It has recently been shown that the 1T phase of some JTMCs is not dynamically stable without biaxial strain \cite{ku2023ab}, even when it is stable compared to 2H.  In addition to the dynamical stability of phonons according to DFPT, as shown in Figures~\ref{fig:elph-tixh}, we investigate the anharmonic thermal effects and transition from 2H to 1T  using \textit{ab initio} molecular dynamics at room temperature (293 K) and below (200 K and 150 K).
    Here, we are interested in energy barriers and anharmonicity, for which a small supercell comprising just four unit cells is sufficient to sample all unstable phonons at K and M.
    The Born-Oppenheimer Hamiltonian energy is shown in Figures~\ref{fig:MD-analysis}. At room temperature, we found that  TiSH, ZrSH, and HfSH did not transform within 4ps, as shown in Figures~\ref{fig:MD-analysis} (a, c, e). At higher temperatures, a transformation occurs as indicated by the rapid drop in the energy, (Figures~\ref{fig:MD-analysis} a, c, e). These different phase-transition temperatures correlate with the energy barrier ($E_{B}$), as shown in Figures~\ref{fig:energy-barrier}, which is higher for ZrSH and HfSH than for TiSH.
    
    For the hexagonal 2H Janus transition metal selenium hydrides, we observe a phase transformation in TiSeH, ZrSeH, and HfSeH, as shown in Figures~\ref{fig:MD-analysis} (b, d, f). At room temperature, TiSeH  all four hydrogens move from 2H to 1T sites after 4.0 ps. ZrSeH partially undergoes a phase transition in which two hydrogens move from the 2H position to the 1T position after 4.0 ps. HfSeH also partially undergoes a phase transition, with one hydrogen moving from the 2H position to the 1T position after 4.0 ps. The mixing between 2H and 1T sites at finite temperature can be explained by additional configuration entropy, leading to a third, hydrogen-disordered phase, as outlined in the appendix. Nevertheless, it is clear that the disordering temperature is above room temperature, and at least an order of magnitude higher than the superconducting T$_c$. It means that ordered JTMCs could be grown at room temperature and cooled to their superconducting state. At a lower temperature of 200 K, the phase transition did not occur in the case of ZrSeH, as indicated by the stable orange line in Figure~\ref{fig:MD-analysis} (d), and is less pronounced in TiSeH and HfSeH, which did not transform at 150 K, as shown by the stable green lines in Figures~\ref{fig:MD-analysis} (b, f). 
    
    The MD behavior is similar in sulfur compounds, but higher temperature was require to initiate the transition.  This is expected because the energy barrier is higher for than the selenium counterparts (Figures~\ref{fig:energy-barrier}).

    The MD demonstrates that, transformation between 2H and 1T is not instantaneous, but can occur on a picosecond timescale.  We would therefore expect the 1T phase to be observed at low temperature, except when 2H can stabilized by substrates. Moreover, 
     disorder and partial occupation of the 2H and 1T may be favoured at elevated temperatures because of the small energy difference between the two phases. Experimentally, the very first observation of a transition metal chalcogenide hydride was shown to be the hexagonal 2H phase of MoSH \cite{wan2021synthesis}, even though calculations show a favorable energy phase of 1T \cite{lu2017janus}, which itself is not dynamically stable \cite{ku2023ab}. For our JTMCs, all essential information is summarized in Table~\ref{tab:energy-structure}.

        \begin{table}[h!]
    \centering
	   \begin{tabular}{cccccc}
        \hline
	   JTMCs & 2H-$\omega_{q}$ & 1T-$\omega_{q}$ & $\Delta E_{GS}$ (eV) & $\Delta E_{B}$ (eV) & T$_{pt} (K)$ \\
	   \hline
	   TiSH & stable & stable & 0.3088 & 0.1312 & 400 \\
	   TiSeH & stable & stable & 0.3248 & 0.1133 & 200 \\
	   TiTeH & unstable & stable & - & - & - \\
	   ZrSH & stable & unstable & - & 0.1943 & 400 \\
	   ZrSeH  & stable & stable & 0.2296 & 0.1843 & 293 \\
          ZrTeH & unstable & stable & - & - & - \\ 
          HfSH & stable & unstable & - & 0.1877 & 500 \\
          HfSeH & stable & unstable & - & 0.1705 & 200\\
          HfTeH & unstable & unstable & - & - & - \\
        \hline
	   \end{tabular}
    \caption{The table shows phonon stability from phonon frequencies from irreducible wavevector $q$ of phonon in the Brillouin zone from DFPT ($\omega_{q}$), ground state energy (E$_{GS}$ (eV)), energy barrier $E_B$ and the lowest temperature where the phase transition (T$_{pt}$) from 2H to 1T was observed}
	\label{tab:energy-structure}
	\end{table}

	\subsection{Phonon-mediated superconductivity} 
    We investigated the phonon-mediated superconductivity of our JTMCs. The calculated characteristics of electron-phonon coupling, required to parameterize the isotropic Eliashberg spectral function ($\alpha^2 F(\omega)$) and the frequency-dependent electron-phonon coupling ($\lambda(\omega)$), are shown in Fig.\ref{fig:elph-tixh} (a-f) for the 2H phase, and Fig.\ref{fig:1telph-tixh} (a-e) for the 1T phase. For the 2H phase, it appears that most of the contribution to the electron-phonon coupling arises from phonons in the region near the $\Gamma$ point. This is demonstrated in both the phonon dispersion and the 2D contour of electron-phonon coupling in the Brillouin zone. The main contributions to the isotropic Eliashberg spectral function ($\alpha^2 F(\omega)$) and the energy-dependent electron-phonon coupling ($\lambda(\omega)$) come from these acoustic phonon modes and the lower optical modes 4 and 5. 
    
    By contrast, for the 1T phase, the main contribution of electron-phonon coupling results from the softening of the acoustic flexural phonon mode between the $\Gamma$ and $K$ points as shown in Figures~\ref{fig:1telph-tixh}, exactly where we observe phonon instability in the unstable materials.

    \begin{table}[h!]
    \centering
	   \begin{tabular}{cccccccc}
        \hline
	   Types & $\lambda$ & $\omega_{\text{log}}$(meV) & $\omega_{2}$(meV) & $T_{c}$(K) (AD)& $T_{c}$(K) (gap) & $2\Delta_0/k_B T_c$ & Ref.\cite{ul2024superconductivity,li2024machine}\\
	   \hline
	   2H-TiSH & 1.11 & 14.99 & 25.07 & 15.31 & 26.5 & 4.3 & 9.24\\
	   2H-TiSeH & 0.96 & 15.06 & 26.39 & 11.51 & 21.3 & 4.3 &13.09 \\
	   2H-ZrSH & 2.68 & 5.47 & 15.83 & 15.09 & 25.2 & 4.8 & - \\
	   2H-ZrSeH & 1.48 & 8.05 & 18.91 & 12.08 & 20.1 & 4.5 & - \\
	   2H-HfSH & 2.37 & 5.64 & 16.89 & 13.83 & 23.5 & 4.8 & - \\
          2H-HfSeH & 1.67 & 7.44 & 19.92 & 12.82 & 19.5 & 4.3 & -\\ 
          \hline
	   1T-TiSH & 1.67 & 15.91 & 21.22 & 23.11 & 35.4 & 4.8 & \\
	   1T-TiSeH & 1.11 & 14.09 & 19.12 & 14.36 & 20.6 & 4.2 & 30.2\\
	   1T-TiTeH & 1.08 & 10.73 & 20.36 & 10.62 & 15.0 & 4.0 & 13.31\\
	   1T-ZrSeH & 2.44 & 7.58 & 13.16 & 19.04 & 22.4 & 5.2 & - \\
	   1T-ZrTeH & 1.18 & 9.85 & 17.52 & 11.12 & 11.7 & 4.2 & 9.04\\
        \hline
	   \end{tabular}
    	\caption{The table shows the superconducting properties, including the electron-phonon coupling constant $\lambda$, the logarithmic average of the phonon energy $\omega_{\text{log}}$, the square average of the phonon energy $\omega_{2}$, the critical superconducting temperature $T_{c}$ based on the Allen-Dynes formula, the critical superconducting temperature $T_{c}$ based on the closing of the superconducting gap, and the superconducting gap and temperature ratio $2\Delta_0/k_B T_c$. Recent results published since we completed our calculations are given in the final column}
	\label{tab:tc-superconducting}
	\end{table}

    At this stage, we summarize the trends of $\lambda$ and $\omega_{\text{log}}$, as follows; 
    \begin{itemize}
        \item For fixed M, $\lambda$ decreases as X = S $\rightarrow$ Se $\rightarrow$ Te.
    \item For fixed X, $\lambda$ increases as M = Ti $\rightarrow$ Zr $\rightarrow$ Hf, except for HfSH. \item $\lambda$ of the same compound is higher in 1T than in 2H. \item For  fixed X, $\omega_{\text{log}}$ decreases as M = Ti $\rightarrow$ Zr $\rightarrow$ Hf, except for ZrSH. 
    \item For  fixed M, in the 1T phase, $\omega_{\text{log}}$ decreases as X = S $\rightarrow$ Se $\rightarrow$ Te. \item Conversely,  for a fixed M in the 2H phase, $\omega_{\text{log}}$ increases as X = S $\rightarrow$ Se.

\end{itemize}
    From these electron-phonon couplings, we estimate the critical superconducting temperature ($T_{c}$) using the Allen-Dynes semi-empirical formula. Additionally, we also compute $T_{c}$ from the superconducting gap, which is obtained from numerical simulations of the anisotropic gap equations using the Migdal-Eliashberg equations. We report the critical temperature obtained from two methods: the value given by the Allen-Dynes formula (AD) and the value for closing superconducting gap. These can be taken as upper and lower bounds and give some indication for the uncertainty in our estimation of $T_c$.
    

    We summarize the trends of $T_{c}$, as follows; 
    \begin{itemize}
        \item In the 2H phase, for a fixed M, $T_{c}$ decreases as X = S $\rightarrow$ Se.
        \item For X = S, $T_{c}$ decreases as M = Ti $\rightarrow$ Zr $\rightarrow$ Hf.
        \item For X = Se, $T_{c}$ increases as M = Ti $\rightarrow$ Zr $\rightarrow$ Hf.
        \item In the 1T phase, for a fixed M, $T_{c}$ decreases as X = S $\rightarrow$ Se $\rightarrow$ Te.
        \item For a fixed X, $T_{c}$ increases as M = Ti $\rightarrow$ Zr.
        \item $T_{c}$ of the same compound is higher in the 1T phase than in the 2H phase.
    \end{itemize}
    
     We also computed the superconducting gap $\Delta_k$ on the Fermi surface by solving the anisotropic Migdal-Eliashberg equations. The temperature-dependent energy distribution of the gap is illustrated in Figures~\ref{fig:elph-tixh}. At the low-temperature limit ($T \approx 0$ K). For 2H phase, the superconducting gap values for TiSH and TiSeH are approximately $\Delta_0 \approx 4.9$ meV and $\Delta_0 \approx 3.9$ meV, respectively. For ZrSH and ZrSeH, the superconducting gap values are $\Delta_0 \approx 5.2$ meV and $\Delta_0 \approx 3.9$ meV, respectively. For HfSH and HfSeH, the superconducting gap values are $\Delta_0 \approx 4.9$ meV and $\Delta_0 \approx 3.8$ meV, respectively. For 1T-phase, the superconducting gap values for TiSH, TiSeH, and TiTeH are $\Delta_0 \approx 7.3$ meV, $\Delta_0 \approx 3.7$ meV, and $\Delta_0 \approx 2.6$ meV, respectively. For ZrSeH, and ZrTeH, $\Delta_0 \approx 5.0$ meV and $\Delta_0 \approx 2.1$ meV, respectively.
    
    From the superconducting temperature where the superconducting gaps vanish, we also determine the corresponding ratios $2\Delta_0/k_B T_c$. As shown in Table \ref{tab:tc-superconducting}, these fall in the range 4.5$\pm 0.5$.

\section{Methods}\label{sec11}

 	The computation was performed based on the density functional theory (DFT), as implemented in QUANTUM ESPRESSO (QE) \cite{giannozzi2009quantum,giannozzi2017advanced}. The crystal structures were created using VESTA \cite{momma2011vesta} with the trigonal space group of P3m1 (No.156). The optimization of the crystal structures was achieved by using the Broyden–Fletcher–Goldfarb–Shanno algorithm (BFGS) method \cite{BFGS,liu1989limited}, and by fully relaxing the crystal structures with the force threshold of $10^{-5}$ eV/\AA. The vacuum thickness was set to be 20 \AA\ with the Coulomb truncation along the z-axis \cite{sohier2017density,sohier2017breakdown} to obtain free-standing 2D-JTMD hydride MLs. The optimized norm-conserving Vanderbilt (ONCV) pseudopotentials \cite{hamann2013optimized,schlipf2015optimization} and the generalized gradient energy functional of the Perdew–Burke–Ernzerhof (GGA-PBE) \cite{perdew1996generalized} were used for the exchange-correlation energy functional with the wavefunction and charge density cutoffs of 80 Ry and 320 Ry, respectively. To sample the reciprocal space of the Brillouin zone for high-quality optimization, we implemented a Monkhorst-Pack grid k-mesh \cite{monkhorst1976special} of 48$\times$48$\times$1 k-point grids with Marzari-Vanderbilt-DeVita-Payne cold smearing of 0.02 Ry on the Fermi surface \cite{marzari1999thermal}. For the electronic structure calculations, we used 24$\times$24$\times$1 k-point grids for self-consistent calculations and implemented the optimized tetrahedral method for the calculation of non-self-consistency \cite{kawamura2014improved}, for the calculation of electronic density of states along with the Fermi surface visualized by XCRYSDEN \cite{kokalj2003computer}. The dynamical matrices were calculated using the density functional perturbation theory (DFPT) with dense k-point grids of of 24$\times$24$\times$1 k-point grids and 12$\times$12$\times$1 q-point grids.

We utilize the electron-phonon Wannier-Fourier interpolation method \cite{giustino2017electron,giustino2007electron} implemented in the EPW package \cite{noffsinger2010epw,ponce2016epw} to compute superconducting, enabling accurate calculations of electron-phonon coupling ($\lambda$) and critical temperature (T$_{c}$). This approach also facilitates the analysis of anisotropic Migdal-Eliashberg theory \cite{margine2013anisotropic} by solving the two nonlinear
coupled anisotropic Migdal-Eliashberg equations, 
    \begin{align}
        Z_{nk}(i\omega_{j}) & = 1+\frac{\pi T}{N({\varepsilon_{F})\omega_{j}}}\sum_{mk'j'}\frac{\omega_{j'}}{\sqrt{\omega^{2}_{j'}+\Delta^{2}_{mk'}(i\omega_{j'})}}, \\
        Z_{nk}(i\omega_{j})\Delta_{mk}(i\omega_{j}) & = \frac{\pi T}{N(\varepsilon_{F})} \sum_{mk'j'} \frac{\Delta_{mk'}(i\omega_{j'})}{\sqrt{\omega^{2}_{j'}+\Delta^{2}_{mk'}(i\omega_{j'})}}[\lambda(nk,mk',\omega_j -\omega_{j'})-\mu^*]\delta(\epsilon_{mk'}-\varepsilon_F),
    \end{align}
    self-consistently along the imaginary axis at the fermion Matsubara frequencies $\omega_j = (2j+1)\pi T$. For these computations, we employ the QUANTUM ESPRESSO package with the same computational setting as mentioned above followed by the Wannier-Fourier interpolation to k- and q-point grids of 120$\times$120$\times$1 and 60$\times$60$\times$1, respectively. The use of dense grids ensures the convergence of $\lambda$ values, as evidenced by the stability of the corresponding $\alpha^{2} F(\omega)$ and $\lambda(\omega)$ even with increasing k- and q-point grid densities. The Fermi surface thickness was set to 0.55 eV, with the Matsubara frequency cutoff at 1.0 eV. The Dirac $\delta$ functions were broadened using a Gaussian function with widths of 0.1 eV for electrons and 0.5 meV for phonons. Finally, the Morel-Anderson pseudopotential was set to $\mu^* = 0.1$ for practical purposes.

    The superconducting transition temperature (T$_{C}$) was calculated using the semi-empirical Allen-Dynes formula \cite{allen1975transition} as
    \begin{equation}
    T_{c} = f_1 f_2 \frac{\omega_{\text{log}}}{1.20} \exp\left(-\frac{1.04(1+\lambda)}{\lambda-\mu^{*}(1+0.62\lambda)}\right)
\end{equation}
where the electron-phonon coupling constant $\lambda$ can be calculated from the Eliashberg spectral function by using 
\begin{equation}
    \lambda = 2\int^{\omega}_{0} d\Omega \left(\frac{\alpha^{2}F(\Omega)}{\Omega}\right),
\end{equation}
and the logarithmic average phonon energy can be computed by using 
\begin{equation}
    \omega_{\text{log}} = \exp\left(\frac{2}{\lambda}\int^{\infty}_{0} d\Omega \text{log}(\Omega) \left(\frac{\alpha^{2}F(\Omega)}{\Omega}\right)\right).
\end{equation}
with
    \begin{equation}
        f_1 f_2 = \left(1+\left(\frac{\lambda}{2.46(1+3.8\mu^{*})}\right)\right)^{1/3}\times\left(1+\frac{\lambda^2 (\frac{\omega_2}{\omega_{\text{log}}}-1)}{\lambda^2 +3.31(1+6.3\mu^*)^2}\right).
    \end{equation} 
This $f_1 f_2$ correction factor will be used when the electron-phonon coupling is typically larger than 1.0. The mean-square frequency ($\omega_2$) is given by
\begin{equation}
    \omega_2 = \sqrt{\frac{2}{\lambda}\int_{0}^{\omega_{\text{max}}}\alpha^2 F(\omega)\omega d\omega}.
\end{equation}
  
\section{Discussion and Conclusions}\label{sec13}

	\section{Conclusion}
   We have systematically investigated 2D  Janus IV-B TMDs hydrides (2H-MXH), with $M=\text{Ti, Zr, Hf}$ and $X=\text{S, Se, Te}$ in both 1T and 2H forms. The electronic structures of the  exhibit many similarities in their electronic band structures and Fermi Surface. These JTMCs display metallic behavior due to a single band crossing the Fermi level. The electronic states at the Fermi level are dominated by the $d$ orbitals of transition metal. A high density of states at the Fermi Energy is observed. This is often indicative of low temperature instability to open a pseudogap, which manifests itself as superconductivity or phonon instability: both are observed for different materials in the JTMC-hydride family. Our investigation into the magnetic phases of JTMCs reveals that the ferromagnetic phase is energetically favorable for the 2H phases of TiSH, TiSeH, ZrSH, and ZrSeH, while the non-magnetic phase is most favorable for the 2H phases of HfSH and HfSeH. For the 1T phases, TiSH, TiSeH, TiTeH, ZrSeH, and ZrTeH also favor a ferromagnetic phase. These findings suggest that the predicted T$_{c}$ would be suppressed for those that favor a ferromagnetic phase in this study and other studies \cite{ul2024superconductivity,li2024machine}. The lower energies of the magnetic phase, as shown in Table~\ref{tab:phase-energies}, are about 10-30 meV compared to the normal metallic phase. However, the potential Cooper pair instability in the metallic state has not been considered, which could further reduce the energy of the non-magnetic metallic phase.
   
   The structural stability is evident in the positive definite spectrum of the  phonon density of states. Furthermore, the predicted differences in frequencies between sulfur, selenium and tellurium appear mainly due to their different masses.  The thermal effect of JTMCs is also investigated using ab initio molecular dynamics. The sulfur-based compounds TiSH, ZrSH, and HfSH did not transform to 1T at room temperature, requiring  400 K, 500 K, and 500 K, respectively and then transforming to disordered mixture of 2H and 1T. However,  selenium-based  TiSeH, ZrSeH, and HfSeH did transform at room temperature.  The 1T structure always has lower energy than 2H, although, the stability of the 2H phase can be enhanced by introducing additional strain, equivalent to substrate stabilization. This strain might also lead to the coexistence of 2H and 1T phases due to the small energy difference between them, as indicated by our calculations. The practical importance of substrates is emphasized by the initial discovery of a transition metal chalcogenide hydride being the 2H phase of MoSH, despite the existence of a lower energy 1T phase that is not dynamically stable.

    For electron-phonon coupling, the most important phonon modes are in the region near the $\Gamma$ point and are related to flexural acoustic modes. 
    These same modes are implicated in the phonon instability of some compounds.
    We estimate the critical superconducting temperature ($T_{c}$) using the Allen-Dynes semi-empirical formula. Additionally, we compute $T_{c}$ from the superconducting gap obtained from numerical simulations of the anisotropic gap equations of the Migdal-Eliashberg equations. We report the critical temperature as a range between the Allen-Dynes result and the closing superconducting gap result, representing the lower and upper limits, respectively.
 
  We also determine the superconducting gap to the temperature ratio $2\Delta_0/k_B T_c$. 
    These $2\Delta_0/k_B T_c$ values are systematically  higher than the ideal weak-coupling (BCS) value of 3.53.  This is  due to the significant electron-phonon coupling ($\lambda$), indicative of strong phonon coupling in these JTMC hydrides. Typical T$_c$ for these 2D JTMC hydrides lies in the range 10-30K, an order of magnitude higher than graphene and MoS$_2$.  Their 2D character suggests that superconductivity could be further enhanced by twisted multilayers.
    

\backmatter





\bmhead{Acknowledgements}
	This research project is supported by the Second Century Fund (C2F), Chulalongkorn University. We acknowledge the supporting computing infrastructure provided by NSTDA, CU, CUAASC, NSRF via PMUB [B05F650021, B37G660013] (Thailand). URL:www.e-science.in.th..  GJA acknowledges funding from the ERC project Hecate. This also work used the Cirrus UK National Tier-2 HPC Service at EPCC (http://www.cirrus.ac.uk) funded by the University of Edinburgh and EPSRC (EP/P020267/1).

    \section*{Data availability statement}
	The data that support the findings of this study are available upon reasonable request from the authors.

    \section*{Conflict of interest}
	The authors have no conflicts of interest to declare. All co-authors have seen and agree with the contents of the manuscript and there is no financial interest to report. We certify that the submission is original work and is not under review at any other publication.
 
    \section*{Appendix: Hydrogen disordering}
The existence of two stable sites for hydrogen opens the possibility that surface may become site disordered. 

We can estimate this from a simple model where the energy of the lattice depends on the on-site atoms and the nearest neighbor bondings only. In our system, there are 2 different sites, i.e. H (for 2H) and T (for 1T) sites. If a hydrogen atom occupies the H site, it contributes the on-site energy $E_H$ = A. If a hydrogen atom occupies the T site, it contributes the on-site energy $E_T$. If we set the 1T phase as a reference, then $E_T$ = 0. This can be written as
    \begin{equation}
        E_{\text{onsite}}=\sum_{i}^{\text{all sites}}A S_{i}^{H}+\sum_{i}^{\text{all sites}}(E_T = 0)*S_{j}^{T},\label{eq:A}
    \end{equation}
    where  $S_i^H$ is the occupancy of the site H index i, which can be 1 if it is occupied or 0 otherwise, and $S_j^T$ is the occupancy of site T index j, which can be 1 if it is occupied or 0 otherwise. However, we set the energy of the 1T phase as reference. Thus, the second term vanishes. 

    For the nearest neighbor bonding at the site H index i, we can model it as 
    \begin{equation}
        E_{nns,i}=\sum_{j}^{\text{all nns}}B S_{i}^{H} S_{j}^{T}, \label{eq:B}
    \end{equation}
    where B is the repulsive energy from having two adjacent hydrogens. We define $S_i^H = 1$ if the H site occupies by a hydrogen atom, or 0 otherwise, and $S_j^T = 1$ if the nearest neighboring T site occupies by a hydrogen atom, or 0 otherwise. In the P3m1 structure, there are precisely 3 nearest neighbors (nns). Then the total energy can be written as
    \begin{equation}
        E_{\text{total}}=\sum_{i}^{\text{all sites}}A S_{i}^{H}+\sum_{i}^{\text{all sites}}\sum_{j}^{\text{all nns}}B S_{i}^{H}S_{j}^{T} \label{eq:AB}
    \end{equation}
    In order to gain further understanding of this model, and estimate the parameters from DFT data, let us give some special examples as follows; 
    
\begin{figure}[h]
		\centering  
    	\includegraphics[width=4cm]{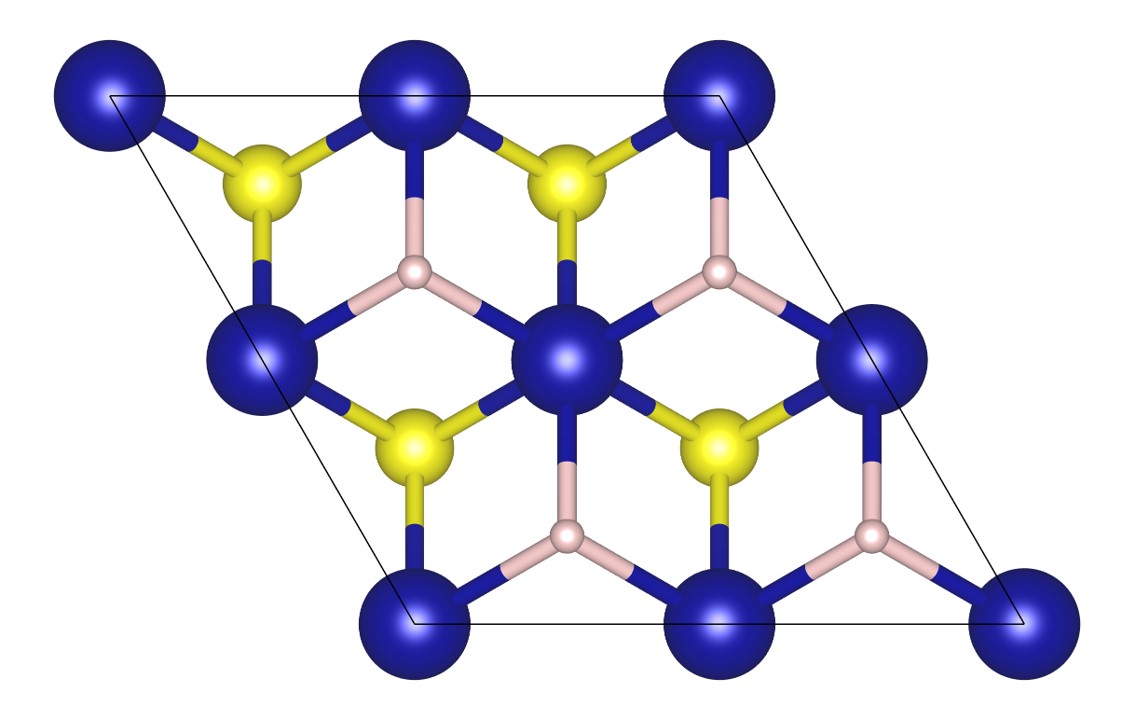}
		\caption{shows the 1T phase, where all T sites are occupied, but all H sites are empty, hence denoted as 0H4T.}
		\label{fig:figureA-1}
	\end{figure}

    1. The supercell of the 1T phase is shown in Figures~\ref{fig:figureA-1}. It can be readily seen that all T sites are occupied, but all H sites are empty, hence this configuration is denoted as 0H4T. This means that $S_i^H = 0$ for all i. Therefore, the total energy is equal to zero.

\begin{figure}[h]
		\centering  
    	\includegraphics[width=4cm]{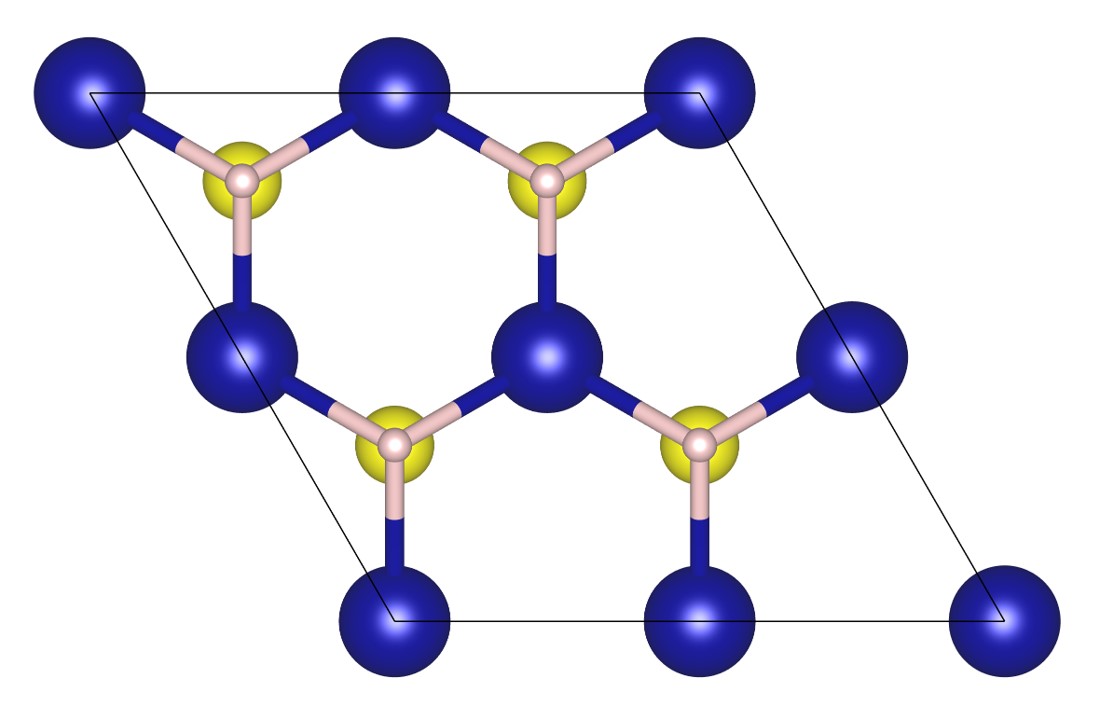}
		\caption{shows the 2H phase, where the T sites are all empty, but all H sites are occupied. In the 2X2 supercell, there are 4 of the H sites. Hence, we denote this configuration as 4H0T.}
		\label{fig:figureA-2}
	\end{figure}

    2. The 2H phase is shown in Figures~\ref{fig:figureA-2}. It can be readily seen that all T sites are empty. This means that $S_j^T = 0$ for all j. In contrast, all H sites are occupied. This means that $S_i^H = 1$ for all i. As a consequence, the cross terms $S_i^H$ $S_j^T$ are all zero. In the 2$\times$2 supercell, there are 4 of the H sites. Hence we denote this configuration as 4H0T. Therefore, the total energy is equal to 4A. From the DFT, the energy of this supercell is equal to 0.132 eV, relative to the 1T phase. Equation \ref{eq:AB} then gives A = 0.033 eV.

\begin{figure}[h]
		\centering  
    	\includegraphics[width=6cm]{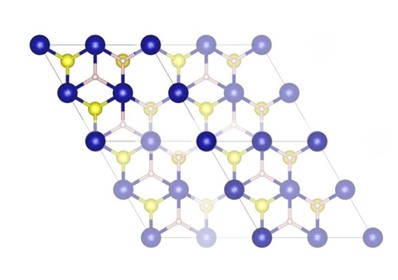}
		\caption{shows a configuration, where 2 out of 4 T sites are occupied, and also 2 of the 4 H sites. Hence, we denote this configuration as 2H2T. }
		\label{fig:figureA-3}
	\end{figure}

    3. In the 2x2 supercell, there are 4 H sites and 4 T sites, but we have only 4 hydrogen atoms. If we put 2 hydrogen atoms into the 4H sites, there will be 36 possible configurations. However, some of these configurations are symmetry-equivalent. Thus, we  study a few special examples. Let us choose a special example as shown in Figure \ref{fig:figureA-3}. It shows an extended supercell which helps us count the exact number of nearest neighbors, including the periodic neighbors. From this figure, we obtain

    \begin{equation} E_{\text{total}}^{2H2T}=\sum_{i}^{\text{all sites}}A S_{i}^{H}+\sum_{i}^{\text{all sites}}\sum_{j}^{\text{all nns}}B S_{i}^{H}S_{j}^{T}=2A+2B
    \end{equation}
    From DFT, the relative energy of this configuration is 0.114589 eV. Hence, we can use our value for A to solve eq.\ref{eq:AB} for B which is equal to 0.024 eV.
    
\begin{figure}[h]
		\centering  
    	\includegraphics[width=8cm]{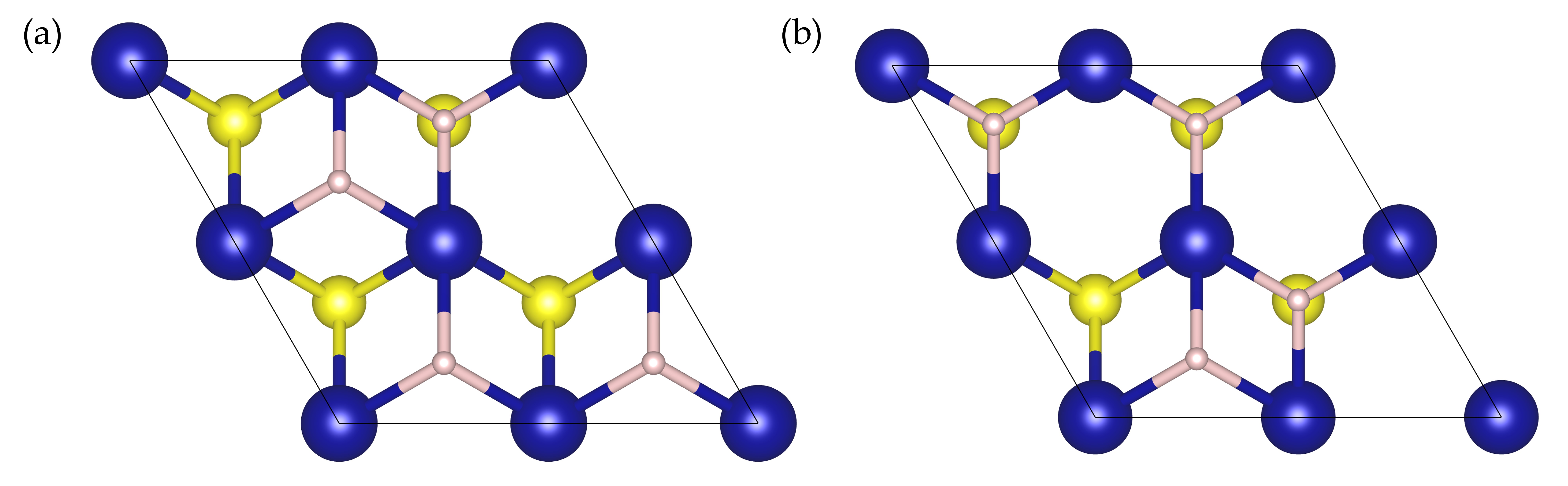}
		\caption{supercells showing (a) 1H3T and (b) 3H1T configurations.}
		\label{fig:figureA-5}
	\end{figure}

    Now, we can test the predictive power of this model. 
    Let us compare our model with DFT results of other configurations, shown in Figure \ref{fig:figureA-5}. From DFT, the 1H3T configuration has energy of 0.0797 eV, compared with the model
    \begin{equation}
        E_{\text{total}}^{1H3T} = A+2B = 0.081 eV.
    \end{equation}
    
    From DFT, the 3H1T configuration in Figure \ref{fig:figureA-5} has energy of 0.147183 eV, comparing with the model prediction 
    \begin{equation}
        E_{\text{total}}^{3H1T} = 3A+2B = 0.147eV.
    \end{equation}
    
   Deviations in energy from equation \ref{eq:AB}  could come from the contribution of the next nearest neighbors, three body terms or also some elastic energy due to supercell relaxation.  However, the accurate reproduction of the configurations not used in fitting show that these effects are negligible.





Finally, we can estimate temperature effect on ordering by considering entropy in the mean-field approximation by comparing the fully ordered 1T phase (the reference state) with the fully disordered phase 
If we assume that the vibration and electronic entropy would nearly cancel between the 1T and 2H phases, the remaining entropy is the configurational entropy, i.e.
    \begin{equation}
        S=-k_{B} \{c \ln c + (1-c) \ln (1-c)\},
    \end{equation}
    where c is the concentration of the H sites relative to the number of hydrogen atoms in the supercell. The fully disordered state has c=0.5, or $S^{disorder} = 0.693k_B$ per site, or twice that per hydrogen atom.   In the mean field case the relative free energy of the disordered phase is:
    $0.5A+1.5B = 0.0525$eV/H atom, because each atom has a 50\% chance of being on an H site, and each of its three neighbouring sites has a 50\% chance to be occupied.   
    
    This is already a relative energy with respect to the 1T phase, so 
    We can estimate the disordering temperature as the zero of relative free energy for the disordered phase, i.e.
 \[ F = 0 = 0.05255-2\times 0.693k_BT \]
    which yields 440K.
 This appendix only outlines the simplest theory of the configuration entropy and for one material only.  Nevertheless, it is clear that the disordering temperature is above room temperature, and at least an order of magnitude higher than the superconducting T$_c$.  It means that ordered JTMCs could be grown at room temperature and cooled to their superconducting state. 

\bibliographystyle{}
\bibliography{sn-bibliography}
\end{document}